\documentstyle[aps,prd]{revtex}

\newcommand{\rf}[1]{(\ref{eq:#1})}

\begin{document}
\title{ Brane-Worlds and their Deformations} 
\author{ M. D. Maia\\
Universidade de Bras\'{\i}lia, Instituto  de F\'{\i}sica \\ 
 Bras\'{\i}lia. D.F. 7919-970\\{maia@fis.unb.br} \\
and\\
Edmundo M. Monte\\
Universidade Federal da Para\'{\i}ba\\
Departamento de Matem\'atica\\
Campina Grande, Pb. 58100-000\\{edmundo@fis.ufpb.br} }

\widetext

\maketitle

\flushright{hep-th 0003196\\
{pacs 11.10.Kk, 04.50.+h, 04.60.+n}
\vspace{1cm}

\begin{abstract}
A geometric theory of   brane-worlds with large  or non-compact
extra dimensions is presented. It is shown that coordinate gauge
independent perturbations of the brane-world correspond to the
Einstein-Hilbert dynamics  derived from the  embeddings of the
brane-world. 
The quantum  states of  a perturbation are described by
Schr\"odinger's equation  with respect to the extra dimensions and
the deformation Hamiltonian.  
A gauge potential  with  confined components is  derived from the
differentiable structure of the brane-world.  
\end{abstract}

\flushleft

\twocolumn
\narrowtext

\section{ Introduction}

Recent theoretical  and  phenomenological  arguments  have suggested
a  unification model,  with   large  extra dimensions at TeV scale, including quantum gravity.
Standard model gauge  fields  and  ordinary matter  remain
confined to the four dimensional  space-time, but  the quantum
gravitational field  propagates in a region  of
the high dimensional space  \cite{Arkani:1,Randall}.  
  Similar concepts  involving   trapped  gauge  and fermionic  fields in a  fluctuating space-time have appeared repeatedly in
the literature, some of them  based on specific models like in
\cite{Akama,RS,Visser},  while others  are based on  different  blends of  Kaluza-Klein
and strings ideas \cite{RT,MM,Pavsic,Gibbons,Antoniadis,Davidson}.   

To make a   theory,  the brane wolds proposition requires  a description of the   space-time  or  brane-world embedded  in a region of a higher dimensional manifold, capable of   undergo  quantum fluctuations and    compatible with  the  confinement of  gauge  field interactions. The implicit  assumption is that  
the hierarchy  of the  involved interactions  maintein stable  the  differentiable  structure of  the space-time.
The purpose of this paper is to  show that  these   properties  derive from  the  geometry of  Riemannian submanifolds and their  deformations.

We start in the next section reviewing the simpler example  of  hypersurface
deformations.  In section III  it is shown that the perturbative analysis
is equivalent  to a  dynamical process resulting from the Einstein-Hilbert
principle. This is generalized  to  multiple  parameter deformations of
submanifolds in section IV.  The quantum  state  of  a deformation
using Schr\"odinger's  equation  with respect to the  extra dimensions is  defined in section V.  The  topological changes associated with these deformations  are also discussed. The paper ends showing that the   differentiable structure of the brane-world  also implies in the existence a  gauge field structure whose  components remain confined to the brane-world.  

\section{Deformations of a Hypersurface}
The following results extends  Nash's  perturbation theorem of  hypersurfaces
in   Euclidean spaces to
a n-dimensional space-time $V_{n}$ embedded in a $(n+1)$-dimensional  manifold
$V_{(n+1)}$ \cite{Nash}. It is  also similar to the deformation of a
3-dimensional hypersurface in space-time within the context of canonical gravity \cite{Teitelboim}.
The reader who is  already familiar with  these techniques may  jump  to the 
next section where  these deformations are  related to a non constrained 
dynamics.   

The isometric embedding of a background space-time $\bar{V}_{n}$ with metric  $\bar{g}_{ij}$    is  given by
the map  $\bar{\cal X}: \bar{V}_{n}\rightarrow V_{n+1}$ such that\footnote{  
All Greek indices  run from 1 to
$n+1$ in this and the next section and from 1 to $D$   in the  remaining of the
paper. Small case Latin indices $i,j,k...$ run from  1  to  $n$.  An overbar
denotes an object of the background  space-time.   
The covariant derivative with respect to
the metric of  the higher dimensional manifold is denoted by  a semicolon and
$\eta^{\mu}_{;i}=\eta^{\mu}_{;\gamma}\bar{\cal X}^{\gamma}_{,i}$ 
denotes  its projection over  $V_{n}$. 
The  curvatures of the higher dimensional space are distinguished
by a calligraphic  ${\cal R} $ .} 
\[
\bar{\cal X}^{\mu}_{,i}\bar{\cal X}^{\nu}_{,j}{\cal G}_{\mu\nu}=\bar{g}_{ij},\;\;
\bar{\cal X}^{\mu}_{,i}\bar\eta^{\nu}{\cal G}_{\mu\nu}=0,\;\;
\bar{\eta}^{\mu}\bar{\eta}^{\nu}{\cal G}_{\mu\nu}= \varepsilon 
\]
where  ${\cal G}_{\mu\nu}$ denotes the   metric of  $V_{n+1}$ in arbitrary coordinates and  $\bar{\eta}^{\mu}$ denotes the  vector normal to $\bar{V}_{n}$ 
with signature $\varepsilon/|\varepsilon| =\pm1$.
A  geometric deformation  of  $\bar{V}_{n}$
along  a direction $\zeta$  is the subset of  $V_{n+1}$  described by 
the coordinates  ${\cal Z}^{\mu}$ given by
the  perturbation of the embedding vielbein $\bar{\cal X}^{\mu}_{,i}$: 
\begin{eqnarray}
{\cal Z}^{\mu}_{,i}(x^{i},s)&=&\bar{{\cal X}}^{\mu}_{,i}+s\pounds_{\zeta}
\bar{{\cal X}}^{\mu}_{,i} 
\label{eq:Defor1}  \\
\eta & =&\bar{\eta} +s\pounds_{\zeta}\bar{\eta}  
\end{eqnarray}

Since  the  space-time   $\bar{V}_{n}$ is  endowed with a general diffeomorphism  group, it is  always possible to find a coordinate system in
which the above Lie  derivative of the tangent component  vanish. This   undesirable  coordinate gauge  should be  excluded from the deformations.
In the example of a homogeneous elastic membrane, the tangent component of the deformation tension is canceled by  the assumption that  it is homogeneous 
and invariant  under  the rotation group of the tangent space.  The consequence is that the   fundamental modes  of  oscillations  are   described by the deformations of the membrane along the orthogonal direction.
In complete analogy with this example, we take  the fundamental
modes of a deformations as given  by orthogonal deformations of a
hypersurface (also  called pure deformations \cite{Hojman}).
Thus, taking  $\zeta = \eta$   in  \rf{Defor1} we obtain    
\begin{eqnarray}
{\cal Z}^{\mu}_{,i}(x,s) &=&\bar{{\cal X}}^{\mu}_{,i}(x)  +s\eta^{\mu}_{,i}(x).
\label{eq:Zi}\\
\eta & = & \bar{\eta} +s[\bar{\eta},\bar{\eta} ]=\bar{\eta} \label{eq:eta}
\end{eqnarray}  
The  embedding of the deformed hypersurface  $V_{n}$ is given by
\begin{equation}
g_{ij}={\cal Z}^{\mu}_{,i}{\cal Z}^{\nu}_{,j}{\cal G}_{\mu\nu},\;\;  
{\cal Z}^{\mu}_{,i}{\eta}^{\nu}{\cal G}_{\mu\nu}=0,\;\;
{\eta}^{\mu}{\eta}^{\nu}{\cal G}_{\mu\nu}=\varepsilon \label{eq:embeddZ}    
\end{equation}
Denoting by $\bar{k}_{ij}= -{\cal X}^{\mu}_{,i}\bar{\eta}^{\nu}_{;j}{\cal
G}_{\mu\nu}$   the  extrinsic curvature of $\bar{V}_{n}$, and
using \rf{Zi}, the deformed  metric $g_{ij}$  can be  written as 
\begin{equation}
g_{ij}={\cal Z}^{\mu}_{,i}{\cal Z}^{\nu}_{,j}{\cal G}_{\mu\nu}=\bar{g}_{ij}
-2s\bar{k}_{ij} +s^{2}\bar{g}^{mn}\bar{k}_{im}\bar{k}_{jn}. \label{eq:gij1}
\end{equation}
and  the extrinsic curvature  of the deformation $V_{n}$  can
be written as  
\begin{equation}
k_{ij}=-{\cal Z}^{\nu}_{,i}\eta^{\mu}_{;j}{\cal G}_{\mu\nu}= \bar{k}_{ij} -s\bar{g}^{mn}\bar{k}_{im}\bar{k}_{jn} \label{eq:kij}
\end{equation}
Comparing \rf {gij1}  and \rf {kij}, we obtain York's relation,  describing the  metric  evolution  with respect to the  deformation
parameter  $s$. 
\begin{equation}
\frac{d g_{ij}}{d s}=\dot{g}_{ij}=-2 {k}_{ij}. \label{eq:YORK}
\end{equation}

Notice that the inverse of  \rf{gij1}   cannot be calculated
exactly in arbitrary dimensions. To obtain  the   contravariant
version  of  \rf{YORK},
consider the matrix notations ${\bf \bar{ g}}=(\bar{g}_{mn})$, ${\bf
g}=(g_{mn})$  and ${\bf k}=(k_{mn})$. Then the  inverse  metric can be 
given  to any order  $(k)$ of approximation  as
\[
\stackrel{(k);\;}{{\bf g}^{-1}}=\left( \sum_{n=0}^{k} (\bar{\bf g}^{-1}{\bf k})^{n}
\right)^{2}\bar{\bf g}^{-1},\;\;\; \; {\bf g} \stackrel{(k)\;\;}{{\bf
g}^{-1}}\approx 1 +0(s^{k+1}) . 
\]
Using this and  $\dot{g}_{im}  \stackrel{(k)}{g^{mj}} +g_{im}
\stackrel{(k)}{\dot{g}^{mj}}=0$ and  defining 
$\stackrel{(k)}{k^{ij}} =\frac{1}{2} \stackrel{(k)}{\dot{g}^{ij}}$,  
we have
\[
\stackrel{(k)\;\;}{k^{ij}}=
{g^{im}}\;\stackrel{(k)}{g^{jn}}k_{mn} 
\] 
Since this is true for all values of  $(k)$, it follows that
\begin{equation}
k^{ij}= +\frac{1}{2}\dot{g}^{ij} \label{eq:YORKCO}
\end{equation}
Consequently,  the   indices  of the extrinsic curvature 
are lowered and risen $g_{ij}$ and $g^{ij}$ respectively,  in accordance with 
\rf{YORK}:   
\begin{eqnarray}
g_{im}g_{jn}k^{mn} & =& g_{im}g_{jn}\frac{1}{2}\dot{g}^{mn}=\nonumber \\
\frac{1}{2}\frac{d}{ds}(g_{im}g_{jn}g^{mn}) &-&
\frac{1}{2}\dot{g}_{im}g_{jn}g^{mn}-
\frac{1}{2}{g}_{im}\dot{g}_{jn}g^{mn}= k_{ij}.\nonumber
\end{eqnarray}

\section{ Dynamics of   Deformations}

From  \rf{embeddZ} it follows that 
\[
g^{ij}{\cal Z}^{\mu}_{,i}{\cal Z}^{\nu}_{,j}{\cal G}_{\mu\nu}= n\;\; \mbox{and}\;\; g^{ij}{\cal  Z}^{\mu}_{,i}{\cal Z}^{\nu}_{,j}
\eta^{\alpha}{\cal G}_{\mu\alpha} =0
\] 
 Therefore, the quantity $\xi^{\mu\nu}=g^{ij}Z^{\mu}_{,i}Z^{\nu}_{,j}$
cannot be proportional  to  ${\cal G}^{\mu\nu}$. Writing
$\xi^{\mu\nu}={{\cal G}}^{\mu\nu} 
+\Psi^{\mu\nu}$, then $\Psi^{\mu\nu}$
 satisfy the conditions 
\[
{\cal G}_{\mu\nu}\Psi^{\mu\nu}=-1\;\; \mbox{ and}\;\; 
\Psi^{\mu\nu}\eta_{\mu}\eta_{\nu}=-\varepsilon
\]
 The  solution of   these  equations,  compatible with  \rf{embeddZ}  is 
$ \Psi^{\mu\nu}=-\varepsilon\eta^{\mu}\eta^{\nu}/\varepsilon $, so that
\begin{equation}
g^{ij}{\cal Z}^{\mu}_{,i}{\cal Z}^{\nu}_{,j}  ={\cal G}^{\mu\nu}
-\frac{1}{\varepsilon}\eta^{\mu}\eta^{\nu}  \label{eq:INV}
\end{equation}

The stability of the  brane-world   with respect  to the  deformation means that the integrability conditions for  the embedding
 \begin{eqnarray}
R_{ijkl} &=&  \frac{2}{\varepsilon} k_{i[l}k_{j]k}  +
 {\cal R}_{\alpha\beta\gamma\delta}{\cal Z}^{\alpha}_{,i}{\cal 
Z}^{\beta}_{,j}{\cal Z}^{\gamma}_{,k}{\cal Z}^{\delta}_{,l}  \label{eq:G}\\ 
 2\kappa_{i[j; k]}  & =& {\cal R}_{\alpha\beta\gamma\delta}{\cal
Z}^{\alpha}_{,i}\eta^{\beta}{\cal Z}^{\gamma}_{,j}{\cal
Z}^{\delta}_{,k}\label{eq:C} 
\end{eqnarray}
must be satisfied. Using  \rf{INV}, the contractions of  \rf{G} and \rf{C}  give
\begin{eqnarray}
R_{jk}& = &\frac{1}{\varepsilon}(\kappa^{\ell}_{k}\kappa_{\ell j} -h\kappa_{jk})
{-\cal R}_{\beta\gamma} {\cal Z}^{\beta}_{,j}{\cal
Z}^{\gamma}_{,k}\nonumber \\
& -& \frac{1}{\varepsilon} {\cal R}_{\alpha\beta\gamma\delta}\eta^{\alpha}\eta^{\delta}
{\cal Z}^{\beta}_{,j}{\cal Z}^{\gamma}_{,k} \label{eq:Rij}\\
R &= & \frac{1}{\varepsilon}(\kappa^{2} -h^{2})  +{\cal R} -\frac{2}{\varepsilon}
{\cal R}_{\alpha\beta}\eta^{\alpha}\eta^{\beta} \label{eq:R}
\end{eqnarray}
and  
\begin{equation}
\kappa^{k}_{i;k} -h_{,i} = {\cal R}_{\alpha\beta}
{\cal Z}^{\alpha}_{,i}{\eta}^{\beta}    
\end{equation}
where we have denoted  by $ h=g^{ij}\kappa_{ij}$ the   mean curvature of
$V_{n}$,  and $\kappa^{2}=k^{ij}k_{ij}$.    

The   Einstein-Hilbert Lagrangian for ${\cal G}_{\mu\nu}$
 derived  directly  from  \rf{R}
is\footnote{ For  generality of  signature, we denote 
$\sqrt{\cal G}$ as meaning $\sqrt{g\varepsilon }$,  where
$g=det(g_{ij})$,  when  $V_{n}$ has  Euclidean signature
and  meaning $\sqrt{-g\varepsilon}$ when  $V_{n}$ has Minkowski signature.}.   
\[ 
{\cal L}={\cal R}\sqrt{{\cal G}}=\left[ {R}-
\frac{1}{\varepsilon}(\kappa^{2}-h^{2})+ \frac{2}{\varepsilon}{\cal 
R}_{\alpha\beta}\eta^{\alpha}\eta^{\beta} 
\right]  
\sqrt{\cal G} 
\] 
The  last term in this Lagrangian  may be calculated in the  Gaussian
normal  frame of the deformation, where  $\eta^{\mu}=\delta^{\mu}_{n+1}$:  
\begin{eqnarray}
{\cal R}_{\alpha\beta}{\eta}^{\alpha}{\eta}^{\beta} &= &
\Gamma^{\alpha}_{n+1\alpha,n+1}-\Gamma^{\alpha}_{n+1\,n+1,\alpha} \nonumber\\
&+&\Gamma^{\beta}_{n+1\,\alpha}\Gamma^{\alpha}_{\beta\, n+1} 
-\Gamma^{\alpha}_{n+1\,n+1}\Gamma^{\beta}_{\alpha\beta} 
= \kappa^{2}-\dot{h} \label{eq:HH}
\end{eqnarray}
Here the dot means derivative with respect to  $s$.
Using \rf{YORK}  it also follows that 
\[
\frac{\partial {\cal
R}_{\alpha\beta}\eta^{\alpha}\eta^{\beta}}{\partial k_{ij}}= 2k^{ij}
-\dot{g}^{ij}=0
\] 
Therefore, after removing the  surface term, the  Lagrangian  reduces to the well-known expression
\begin{equation}
{\cal L}=\left[ {R} +
\frac{1}{\varepsilon}(\kappa^{2}+h^{2})\right]  \sqrt{{\cal G}}\label{eq:L} 
\end{equation}
The sequence is  similar to the  standard  $3+1$  space-time decomposition, except that here we are using an internal parameter.
The  momentum canonically conjugated to  ${\cal G}_{\alpha\beta}$,  with respect to the  deformation parameter 
$s$ is defined  by 
\[
p^{\alpha\beta}=\frac{\partial{\cal L}}{\partial
(\dot{\cal G}_{\alpha\beta}) }
\]
Using \rf{L}, the momentum components corresponding to
$g_{ij}$  are      
\begin{equation}
p^{ij} =\frac{-1}{\varepsilon}( k^{ij}+hg^{ij} )\sqrt{{\cal G}} \label{eq:piij}
\end{equation}
Notice that the compatibility between  \rf{YORK}  and  \rf{YORKCO} and the tensor algebra of
$V_{n}$  requires that  $ p_{ij}= -\partial{\cal L} /\partial \dot{g}^{ij}$.

Since  the  deformation  does not
prescribe   the evolution of   ${\cal G}_{i\,n+1}$ and ${\cal G}_{n+1\,n+1}$,
 the  corresponding  momenta  do not  follow  from
York's relation and their values are set as constraints to the deformations: 
\begin{eqnarray}
p^{i\, n+1} & = &  -2\frac{\partial{\cal
R}_{\alpha\beta}\eta^{\alpha}\eta^{\beta} }{\partial \dot{{\cal
G}}_{i\,n+1}}\sqrt{{\cal G}}=0  
\label{eq:DEFCON1},\vspace{3mm}\\ 
p^{n+1\,n+1} & = & -2\frac{\partial {\cal R}_{\alpha\beta}\eta^{\alpha}\eta^{\beta} }
{\partial {\dot{\cal G}}_{n+1\,n+1}}\sqrt{{\cal G}}=0 \label{eq:DEFCON2}
\end{eqnarray}
The constraint  \rf{DEFCON1}  is  equivalent  to the fact that the  deformation  does not
 have tangent  components,  a  condition  already  imposed  to guarantee the coordinate gauge independence. Equation \rf{DEFCON2}
corresponds to the fact that the  evolution of the system is measured by the
parameter $s$ alone,  not depending on the  variations of $g_{n+1\,n+1}$. 

Using the definition $p={\cal G}_{\alpha\beta}p^{\alpha\beta}$,
\rf{DEFCON1} and \rf{DEFCON2},   we obtain the  expressions
\[
k_{ij} =\frac{\varepsilon}{\sqrt{\cal G}} (\frac{p g_{ij}}{n+1}-p_{ij}),
\;  h=\frac{-\varepsilon p}{(n+1)\sqrt{\cal G}}
\]
In addition,  $\kappa^ {2}+h^ {2}= -(\frac {p^ {2}} {n+1} -p^ {ij} p_ {ij})/
{\cal G} $.

 The deformation  Hamiltonian is  obtained from the usual Legendre
transformation 
\begin{eqnarray}
{\cal H}=p^{\alpha\beta}\dot{\cal G}_{\alpha\beta}-{\cal L}  &=&\nonumber\\
-R\sqrt{{\cal G}} &-&\frac{\varepsilon}{{\cal G}}\left( \frac{p^{2}}{n+1} -p_{ij}p^{ij}
\right) \sqrt{{\cal G}}
\label{eq:H}
\end{eqnarray}
and Hamilton's  equations  with respect to  $s$  are
\begin{eqnarray}
\dot{g}^{ij} &=&
\frac{-2\varepsilon}{\sqrt{{\cal G}}}\left (\frac{p}{n+1}g^{ij}- p^{ij} \right)
\sqrt{{\cal G}},
\label{eq:DOTG}\\
\dot{p}^{ij} &=  &
- (R^{ij}-\frac{1}{2}Rg^{ij})\sqrt{-{\cal G}}\nonumber \\
&+& \frac{1}{\sqrt{{\cal
G}}}\left[p p^{ij}- 
2p^{im}p^{j}_{m} 
+\frac{1}{2}(\frac{p^{2}}{2}-p^{mn}p_{mn} ) g^{ij} \right ]
\label{eq:DOTP} 
\end{eqnarray} 
The equation  in $g_{ij}$  coincides with   \rf{YORK}  
expressed in terms of  $p_{ij}$.  It follows that  the  deformation
originally described as a perturbative process is an exact  dynamical process
governed by the Einstein-Hilbert principle. 

Hypersurface  deformations  do not depend on the compactness of the extra dimension. In principle  we could apply it to implement the Randall-Sundrum model \cite{Randall}, taking  $V_{n+1}$   to be a  5-dimensional  AdS manifold. In this  case the   extra dimensional slice  is  determined by the  limits in $s$  imposed by the regularity of  \rf{gij1}. This  will be discussed  in the next sections together with the quantum deformations.

\section{Brane-world   Deformations}
A single  extra degree of freedom is  not  compatible with the full
development of  brane-world theory,  especially in what concerns 
quantum gravity and the spin-statistics theorem \cite{Sorkin},  and  with the 
standard  model of gauge interactions.
 In this section  we will  see that  the  confinement of  a geometric gauge field requires  at  least two  extra dimensions.

A given  background space-time   $\bar{V}_{n}$  may  be locally embedded
into a manifold  $V_{D}$, with sufficiently large dimension $D$ and  with
metric signature $(P,Q)$. The number of extra dimensions  $N=D-n$  
depends on the  geometries of  $\bar{V}_{n}$ and of  $V_{D}$, and  of course
of the  embedding map. 

The best known  examples  are  those of space-times
isometrically embedded in a flat space  $M_{D}$, where the  embedding is
given by  analytic functions  \cite{Janet}. The analyticity simplifies
some embedding results  but it imposes  a
maximum embedding dimension  for all  four-dimensional space-times to be  $D=10$. 
Since it is  not obvious  that the analyticity will hold at TeV  scale
excitations,  we may assume  that the deformed  manifolds
remain (at least) differentiable.  In this case, 
the  limit dimension for flat embeddings rises  to $D=14$,  with a   wide
range of compatible signatures \cite{Greene}. 
However, in  general  $V_{D}$  is not  flat and   its  actual dimension 
and signature depends on the geometry imposed on $V_{D}$. In agreement  with  the Kaluza-Klein  principle,  $V_{D}$
is taken to  be   a  solution of the  higher dimensional Einstein's equations,  with or
without a  positive or negative cosmological term.

The multiparameter deformation is  a straightforward generalization of
\rf{Zi} and  \rf{eta},  extended to  the  $N=D-n$ 
directions orthogonal to the background  space-time: 
\begin{eqnarray}
{\cal Z}^{\mu}_{,i} & = & \bar{X}^{\mu}_{,i}  +  s^{A}\eta^{\mu}_{A,i} \label{eq:embeddZA}\\
\eta^{\mu}_{A} & =& \bar{\eta}^{\mu}_{A} + s^{B}[\bar{\eta}_{A},\bar{\eta}_{B}]
\end{eqnarray}
With the condition that  each  $\eta_{A}$  is  independent of the other, 
the Lie bracket in the last  expression vanishes so that it simplifies to
$\eta^{\mu}_{A}= \bar{\eta}^{\mu}_{A}$. 
Notice however that there is a degree  of  freedom in the choice of the
orthogonal basis $\{ \bar{\eta}_{A}\}$,  given by the  isometry group
of the internal  space  $B_{N}$ tangent to the  normal vectors $\bar{\eta}_{A}$.  This  symmetry will impact on the  brane-worlds gauge structure.

The embedding of  the deformed submanifold is  now given by
 \begin{equation}
{\cal Z}^{\mu}_{,i}{\cal Z}^{\nu}_{,j}{\cal G}_{\mu\nu} =g_{ij},\;
{\cal Z}^{\mu}_{,i}\eta^{\nu}_{A}{\cal G}_{\mu\nu}=g_{iA},\; 
{\eta}^{\mu}_{A}{\eta}^{\nu}_{B}{\cal
G}_{\mu\nu}=g_{AB}   \label{eq:multi}
\end{equation}
where $g_{AB}$ denotes the  metric of the  internal space $B_{N}$ with tangent
vectors $\eta_{A}$.
The cross  terms are    $g_{iA}=s^{M}A_{iMA}$, where  
\begin{equation}
A_{iAB}=\eta^{\mu}_{A,i}\eta^{\nu}_{B} {\cal G}_{\mu\nu}   \label{eq:AiAB}
\end{equation}
which appear only when  we have two or   more extra  dimensions. 

The  extrinsic curvatures are  defined for each  direction $\eta_{A}$ as
\begin{equation}
k_{ijA}=-{\cal Z}^{\mu}_{,i}{\eta}^{\nu}_{A;j}{\cal G}_{\mu\nu} \label{eq:KijA}
\end{equation}
From  \rf{embeddZA}  we obtain
\[
g_{ij}\! =\! \bar{g}_{ij} -2s^{A}\bar{\kappa}_{ijA}
+s^{A}\!s^{B}(\bar{g}^{mn}\bar{\kappa}_{imA}\bar{\kappa}_{jnB} + g^{MN}\! A_{iMA}A_{jNB}\!)
\]
and
\[
\kappa_{ijA} =\bar{\kappa}_{ijA} -
s^{B}(\bar{g}^{mn}\bar{\kappa}_{miA}\bar{\kappa}_{jnB} +g^{MN}{A}_{iMA}A_{jNB}) 
\]   
The derivative of  $g_{ij}$ with respect to  $s^{A}$ gives a generalization of \rf{YORK},  describing the  brane metric  evolution along the extra dimension $\eta_{A}$
\begin{equation}
\frac{\partial g_{ij}}{\partial s^{A}}=-2\kappa_{ijA} \label{eq:YORKG}
\end{equation}
The  mean  curvature  is also defined for each  normal  direction as
$h_{A}=g^{ij}\kappa_{ijA}$  and   its norm   is 
 $h^{2} =g^{AB}h_{A}h_{B}$.
As  in the case of  hypersurfaces,  we apply  the integrability conditions for
the embedding,  required to guarantee that the  deformation is again  an
embedded submanifold:  
\begin{eqnarray}
R_{ijkl}& =& 2g^{MN}\kappa_{i[kM}\kappa_{jl]N} +{\cal R}_{\mu\nu\rho\sigma}{\cal Z}^{\mu}_{,i}
{\cal Z}^{\nu}_{,j}{\cal Z}^{\rho}_{,k}{\cal Z}^{\sigma}_{,l}\nonumber \\
\kappa_{i[jA,k]}& =& g^{MN}A_{[kMA}\kappa_{ij]N} +{\cal R}_{\mu\nu\rho\sigma}
{\cal Z}^{\mu}_{,i} \eta^{\nu}{\cal Z}^{\rho}_{,j}{\cal Z}^{\sigma}_{,k}\label{eq:GCR}\\
2A_{[jAB;k]} &=& - 2g^{MN}A_{[jMA}A_{k]NB} \nonumber \\
 &-& g^{mn}\kappa_{[jmA}\kappa_{k]nB} - {\cal R}_{\mu\nu\rho\sigma}{\cal
Z}^{\rho}_{,j} {\cal Z}^{\sigma}_{,k}\eta^{\nu}_{A}\eta^{\mu}_{B} \nonumber
\end{eqnarray}

To write the Einstein-Hilbert Lagrangian  we start again
with the  inversion of the embedding. Following the same procedure as in
\rf{INV},  we find that
\begin{equation}
g^{ij}{\cal Z}^{\mu}_{,i}{\cal Z}^{\nu}_{,j}  ={\cal G}^{\mu\nu}
-g^{AB}\eta^{\mu}_{A}\eta^{\nu}_{B}  \label{eq:INV1}
\end{equation}
Applying this  in the first   equation  \rf{GCR} we obtain
\begin{eqnarray}
R  &=& (\kappa^{2} -h^{2})  +{\cal R}  -2g^{MN}{\cal R}_{\mu\nu} \eta^{\mu}_{M}\eta^{\nu}_{N}\\
& - & g^{AB}g^{MN}{\cal R}_{\mu\nu\rho\sigma}\eta^{\mu}_{A}\eta^{\sigma}_{B}\eta^{\nu}_{M}
\eta^{\nu}_{N} \nonumber       
\end{eqnarray}
where  we have denoted $\kappa^{2}=\kappa_{ijA}\kappa^{ijA}$.
Using the extended Riemann normal coordinates  defined by the deformed submanifold and $\eta^{\mu}_{A}=\delta^{\mu}_{A}$,  it follows that the last term vanishes and  that
\[
g^{AB}{\cal R}_{\mu\nu}\eta^{\mu}_{A}\eta^{\nu}_{B}  =-g^{AB}\frac{\partial
h_{A}}{\partial s^{B}} +\kappa^{2}
\]
Therefore,
\[
 R ={\cal R} - (\kappa^{2} + h^{2})  -2 g^{AB}\frac{\partial h_{A}}{\partial s^{B}}
\]
After discarding the hypersurface terms, the Einstein Hilbert Lagrangian for
$V_{D}$  becomes remarkably simple 
\begin{equation}
{\cal  L}(g,g_{,A})=  \left[ R  + (\kappa^{2} + h^{2})\right] \sqrt{{\cal G}}
\end{equation}

The  momentum  canonically conjugated to the metric  with respect to the normal direction $\eta_{A} $ is
\[
p^{\alpha\beta}_{( A)} =\frac{\partial {\cal L}}{     \partial \left(
 \frac{ \partial {\cal G}_{\alpha\beta} }{\partial s^{A}}   \right)     }
\]
In particular, using  \rf{YORKG}  we obtain the components
\begin{equation}
p^{ij}_{(A)}=-(\kappa^{ij}_{A} +h_{A}g^{ij})\sqrt{{\cal G}} \label{eq:PijA}
\end{equation}
As  before, the  remaining components are    taken as  momentum constraints
\begin{eqnarray}
p^{iA}_{(B)}&=&0\\
p^{AB}_{(C)}&=&0
\end{eqnarray}
We notice also that the  compatibility of \rf{YORKG} with the tensor algebra of  $V_{n}$ requires  the definition of the contravariant momentum components  to be 
\[
p_{ij(A)}=-\partial{\cal L} /\partial g^{ij}_{,A} =-(\kappa_{ijA} +h_{A}g^{ij})\sqrt{{\cal G}}
\]
Denoting  $p_{A}=g^{ij}p_{ij(A)}$,  $p^{A}=g^{AB}p_{B}$  we obtain  
 $h_{A}=-p_{A}/ (n+1)\sqrt{{\cal G}} $ and  defining the orthogonal momentum
norm   $p^{2} =g^{AB}p_{A}p_{B} $, it follows that
\[
h^{2} = \frac{p^{2}}{(n+1)^{2}{\cal G}},\;\; \kappa^{2}= \frac{1}{{\cal G}}\left(
\frac{-(n+2)}{(n+1)^{2}}p^{2} +p^{ijA}p_{ijA} \right) 
\]
The  Hamiltonian  of the  deformation   is  defined by the  Legendre
transformation 
\begin{eqnarray}
{\cal H} &=&  \sum_{A=n+1}^{D} p^{ij(A)}g_{ij,A}-{\cal L} =\nonumber \\
 &-& R\sqrt{{\cal G}}  -\frac{1}{{\cal G}}\left(\frac{p^{2}}{n+1}
-\sum_{A=n+1}^{D} p_{ij (A)}p^{ij(A)} \right) \label{eq:HG}
\end{eqnarray}
Except for  the signature  and the number of dimensions this is
the same  Hamiltonian  \rf{H}, leading to similar  Hamilton's equations,
one for  each internal index  $(A)$
\[
\frac{d g_{ij}}{d s^{A}} =\frac{\delta {\cal H}}{\delta p^{ij}_{(A)}},\;\;\;
\frac{d p^{ij}_{(A)}}{d s^{A}}=-\frac{\delta {\cal H}}{\delta g_{ij}}
\]
The first of these equations coincide with  \rf{YORKG} and the second equation
expresses the variation of the  extrinsic curvature in terms of  the momentum.
The conclusion  is the  same as before: The deformation of  a brane-world  can
be derived  from  the Einstein-Hilbert  action for  the  metric of  $V_{D}$.
The difference is that now  the internal symmetry  has to be  taken into
account.

\section{Quantum Brane-world  Deformations}
One of the basic requirements of brane-world is that  the extra dimensions  are accessed  by   quantum fluctuations of the space-time geometry. 
The meaning  of quantum  
geometry is not  at all clear as it requires the adaptation of concepts that are typical of  physics to geometry. For  example, in such discipline  we could talk  about an atom (or  quantum) of length, area,  connection and curvature \cite{Ashtekar}. Going further we may  speculate on the quantum topology of a  manifold and its  consequences \cite{Hawking,Balachandra,Dowker}, regardless if this is  a metric topology or not. 

In brane-worlds the metric is  also  a physical field  and 
the quantum geometry comes after the  quantization of the  metric. However, the   quantization of the relativistic gravitational
field as  an isolated system  has proven to be  a difficult subject. In fact, 
the  gravitational field  is a prime example of  a 
constrained canonical system, to which  
 Dirac's standard  procedure for  such systems could be applied.    
As it happens, the requirements  of  diffeomorphism invariance of
the theory implies that  the propagation  of the Poisson's bracket
structure does not produce  the expected results \cite{Teitelboim},
except possibly using preferred frames.  

However,  with brane-worlds  the basic  requirement  is that  of  a quantum  fluctuation of  the   space-time, prior to  any consideration on the  second quantization of  the metric field.
The suggestion comes from the compactification  hypothesis   introduced by O. Klein, to make  Kaluza's theory  consistent  with quantum mechanics. Accordingly,
all functions  are  harmonically expanded in  terms of the internal parameters\cite{Klein}.  Therefore, these   discrete internal modes  would correspond to the   quantum modes   with respect to the internal parameters, as if they were internal times.
In brane-worlds   such  discrete modes are more difficult to realize, firstly because the  internal space is not necessarily compact
and secondly because  of the  hierarchical   distinction  between  the gravitational and  gauge  interactions. Nonetheless,  it is possible to quantize the brane embedding map with respect to the  internal parameters.
In fact,  we have seen that  the  deformations of the space-time submanifold geometry with respect to the internal parameters
 correspond to the deformation Hamiltonian \rf{HG}.
Since the  internal parameters  do not share the same
diffeomorphism of the  space-time coordinates, we may use
Schr\"odinger's picture to describe  the  quantum  wave function of a deformation  along a direction $\eta_{A}$ as a solution of  
\begin{equation}
-i\hbar \frac{ d \Psi_{A} }{ d s^{A}} =\hat{\cal H}\Psi_{A}
\label{eq:SCA}
\end{equation}
where  the operator $\hat{\cal H}$ is  constructed with
\rf{HG}.  
The probability  of  a  deformation being in a state   $\Psi_{A}$ is  given by
  $||\Psi_{A}||^{2}=\int \Psi_{A}^{\dagger}\Psi_{A}dv$,  with the integral
extending over the volume of the deformed region in $V_{D}$.

The superposition of quantum deformations states  defined in   the same region of   $V_{n}$ is given by $\Psi  =\sum \Psi_{A}$.
Classically, this  
superposition correspond to a  deformed submanifold along a  direction 
$ \zeta  = \sum_{A}s^{A}\eta_{A} $, 
with  norm   is   $|\zeta|^{2} =\sum g_{AB}s^{A}s^{B} =s^{2}$.
The classical deformation along the  unit direction    $\eta= \zeta /s$ is given by 
\begin{equation}
{\cal Z}^{\mu}_{,i}  =\bar{\cal X}^{\mu}_{,i} +s\eta^{\mu}_{,i}=\bar{\cal
X}^{\mu}_{,i} +\sum_{A}s^{A}\eta^{\mu}_{A,i} \label{eq:ZMUI}
\end{equation}
Given two deformations $\Psi_{A}$ and $\Psi_{B}$, the  transition probability
between  them is   given  by the  Hilbert  product
$<\Psi_{A},\Psi_{B}>$.
The interpretations of the classical limit of the  manifold quantization
depends on the signature of  $V_{D}$.    For  example, if  $\eta_{A}$ and  $\eta_{B}$ are  both space-like, then  
 $<\Psi_{A},\Psi_{B}>$ corresponds in the classical limit to a space-like  handle.  
On the other hand, if  $\eta_{A}$ and  $\eta_{B}$  have  both time-like signatures 
the classical limit  of  $<\Psi_{A},\Psi_{B}>$  corresponds to a classical loop involving two internal time parameters. 
In this case,  a  deformation along $\eta_{A}$  with  evolution scale  $s^{A}$  has  a transition  to  $\eta_{B}$   
with different  scale $s^{B}$ as if an internal time machine. Finally, if  $\eta_{A}$ and  $\eta_{B}$  have different signatures, then the transition probability  $<\Psi_{A},\Psi_{B}>$  corresponds to a signature change.
An  example is given by the  Kruskal space-time  regarded as  a deformation of the Schwarzschild  space-time, in such a way that the latter becomes geodesically complete. These spaces  are both minimally embedded in  six
dimensional spaces but  with  different  signatures   $(5,1)$ and $(4,2)$ respectively.
Since  the Schwarzschild space-time  is a subset of  Kruskal space-time, we  cannot have  both space-times in the same fixed embedding space.  However, they can be considered as  classical limits of  a quantum deformation of the Schwarzschild space-time with a signature transition at the horizon. 
This  agrees with the  known fact that  those solutions have different topologies.

Equation  \rf{SCA} is constructed with the  Einstein-Hilbert  action  and  therefore it may be  referred to as  a fist  quantization of the  space-time geometry where the  expectation value  of the metric is given  by $<\Psi| \hat{g}_{ij}|\Psi>$. 
The second  quantization of the  metric as  a field
 can be adapted  from many of the current  ideas to the  brane-world configuration. In one example  we may regard the gravitational field as  an effective field theory in $D$ dimensions, where $V_{D}$ is seen as the  
space of  all deformed metrics,  in which the effective Planck mass is taken as the regularization  mass \cite{DeWitt}.

\section{Confined Gauge Fields}

As we have seen, equations  \rf{GCR} are responsible for the  stability of
submanifold structure under classical deformations. Therefore, these equations
should  also say  about the  confinement of  the  gauge interactions. In fact, the  basic  field variables in 
\rf{GCR} are  $g_{ij}$, $\kappa_{ijA}$ and  $A_{iAB}$. The first two  vary
with the deformation and they are related by \rf{YORKG}. On the other hand,
$A_{iAB} =-A_{iBA}$  and
\begin{equation}
A_{iAB}  =\eta^{\mu}_{B;i}\eta^{\nu}_{A}{\cal
G}_{\mu\nu}= \bar{\eta}^{\mu}_{B;i}\bar{\eta}^{\nu}_{A}{\cal G}_{\mu\nu}
=\bar{A}_{iAB} 
\end{equation}
so that $A_{iAB}$  does not  propagate with  $s^{A}$. It was shown elsewhere   the relevant property 
that  $A_{iAB}$   transforms like the components of a gauge  field with respect to  the
group of isometries of  $B_{N}$  \cite{Maia,ME}.
Actually,   $A_{iAB}$  is  a Yang-Mills potential  in  the  Einstein-Yang-Mills
equations for a deformed metric  with respect to  the mentioned
group of isometries.  To see this, consider
 the metric  of  $V_{D}$    written in the
Gaussian normal frame of  the deformed  submanifold,
with separate components 
\begin{eqnarray}
g_{ij} &= &{\cal Z}^{\mu}_{i,}{\cal Z}^{\nu}_{,j}{\cal G}_{\mu\nu}=
\bar{g}_{ij} -2s^{A}\bar{k}_{ijA}\nonumber\\
&+& s^{A}s^{B}\left ( \bar{g}^{mn}\bar{k}_{imA}\bar{k}_{jnB}
+\bar{g}^{MN}A_{iMA}A_{jNB}\right )\nonumber\\ 
g_{iA} & = & {\cal Z}^{\mu}_{,i}\eta^{\nu}_{A}{\cal G}_{\mu\nu} =
s^{M}A_{iMA}\nonumber  \\
g_{AB} & =&\eta^{\mu}_{A}\eta^{\nu}_{B}{\cal G}_{AB} \nonumber 
\end{eqnarray}
or,  after   denoting
\begin{eqnarray}
\tilde{g}_{ij}& =&\bar{g}_{ij} -2s^{A}\bar{k}_{ijA} +s^{A}s^{B}\bar{g}^{MN}
\bar{k}_{imA}\bar{k}_{jnB}  \label{eq:gij}\\
  A_{iA} & = & s^{M}A_{iMA} \label{eq:AiA}
\end{eqnarray}
we may write the  metric of  $V_{D}$ as
\begin{equation}
{\cal G}_{\alpha\beta}=
\left( \matrix{ \tilde{g}_{ij}  + g^{MN}A_{iM}A_{jN}  &   A_{iA} \cr
                           A_{jB} &  g_{AB}}
\right) \label{eq:KK}
\end{equation}
This   has the same appearance as the Kaluza-Klein metric ansatz,  with the exception that the cross terms  are   determined by $A_{iAB}$. As with Kaluza-Klein theory,    the
Einstein-Hilbert  Lagrangian  derived from \rf{KK} is
\begin{equation}
{\cal L } ={\cal R}\sqrt{{\cal G}}  ={R}\sqrt{\varepsilon{g}}
+\frac{1}{4}tr {F}^{2}\sqrt{\varepsilon g} 
+\;\;\mbox{\small eventual extra terms}   \label{eq:EYM}
\end{equation}
where we have denoted  $\epsilon=det(g_{AB})$,
${F}^{2}={F}_{ij}{F}^{ij}$ and  ${F}_{ij}=[D_{i},D_{j}]$ where
$D_{i}=\partial_{i} + A_{i}$, with the connection $A_{i}$   given by
\[
{A}_{i}=s^{M}{A}_{iMA}K^{A}
\]
Here  $\{K^{A}\}$ denotes the Killing basis of the
Lie algebra of the group of isometries  of the metric $g_{AB}$.

The  Einstein-Yang-Mills equations   derived  from the
above Lagrangian provide the dynamics of the confined gauge fields.
In addition it provides the  necessary  relationship  between the  internal parameters and the space-time geometry.  

The range of the internal parameters   $s^{A}$  depends on the  curvature of $\bar{V}_{n}$.  This can be seen  after writing   \rf{gij} as
\[
\tilde{g}_{ij} = \bar{g}^{mn}(\bar{g}_{im} -s^{A}\bar{\kappa}_{imA})(\bar{g}_{jn}  -s^{B}\bar{\kappa}_{jnB})
\]
Since the determinant equation $det(\bar{g}_{ij}-s^{A}\bar{\kappa}_{ij})=0$ 
defines the  manifold  of curvature centers of  $\bar{V}_{n}$,  the values of  $s^{A}$  are  limited by this  manifold,
which is  a singular  and forbidden region for  \rf{KK}. 
Denoting by $\tilde{V}_{n}$ the closest regular manifold to that manifold,
the parameters  $s^{A}$ are  restricted  to the region  of 
the higher dimensional space sandwiched between the background $\bar{V}_{n}$ and  $\tilde{V}_{n}$. This  provides an explanation for the
 thickness hypothesis in the 5-dimensional example in \cite{Randall}.

We would like to end  with a  comment on the   dimension of the brane-world.
The  higher dimensional space  $V_{D}$ may be  thought of as composed of  substructures of
different degrees of complexity, which are  submanifolds  of dimension
$n$, where the value of $n$ can be anything from $0$ to $D$. 
The simplest case $n=0$  corresponds to  point particles. 
With  $n=1$,   we  may
formulate a classical string theory. For  $n \ge 2$  we obtain  surfaces and
in general n-branes. In this last case the  internal space  play a  more significant role.  In fact, \rf{GCR}
are differentiable  equations, which 
can be solved  without appeal to  the analyticity of the embedding functions.
If so, then  the limiting dimension  changes to $D=n(n+3)/2$, so that  $n^{2} +n
-2N\ge 0$.  
Taking the  Standard model gauge group $SU(3)\times SU(2)\times U(1)$ acting on
a seven dimensional projective space,  the nearest integer dimension  is  $ n =4$. 
More appropriately, we should look for the GUT  group  which also 
acts as the
group of isometries of $B_{N}$. The  $SO(10)$ model gives  exactly the
value   $n=4$, suggesting a particular fourteen dimensional  model  
 with signature $(11,3)$  and  with   $SO(1O)$ as   the gauge group.


\begin{thebibliography}{43}

\bibitem{Arkani:1} N. Arkani-Hamed et al,  Phys. Lett. {\underline{B429}, 263 (1998)
 hep-th/9803315;  Phys. Rev. Lett. {\underline{84} 586} (2000),
hep-th/9907209} 
\bibitem{Randall} L. Randall \& R. Sundrum,  Phys. Rev. Lett. {
\underline{83}, 4690 (1998), hep-th/9906064}; { Phys. Rev. Lett.
\underline{83}, 3370 (1999),  hep-th/9905221}
\bibitem{Akama} K. Akama,  {\em Pregeometry}, in Gauge Theory  and Gravitation, Lecture notes in Physics 176 (Springer Verlag (1983))
\bibitem{RS}  V. A. Rubakov \& M. E. Shaposhnikov, Phys. Lett. {\underline{125B}, 136, (1983)}   
\bibitem{Visser} M. Visser,  Phys. Lett.  { \underline{159B}, 22  (1985)},
hep-th/9910093; 
\bibitem{RT}   T. Regge \& C. Teitelboim, C.  {\em Relativity a la  String}
 Proc. $I^{st}$ Marcell Grosmmann Meeting, Trieste { (1975)};
S. Deser  et al, Phys. Rev.  { \underline{D14}, 3301, (1976)}.
\bibitem{MM} M. D.  Maia  \&  W. Mecklemburg, Jour. Math. Phys. {
\underline{25}, 3047 (1984) }; 
 M. D. Maia,  Phys. Rev.D {\underline{31}, 262, and  268 (1985)}
\bibitem{Pavsic}  M. Pavsic,  Nuovo Cimento {  \underline{A108}, 221 (1995)}
\bibitem{Gibbons}  G.W. Gibbons \& D. L. Wiltshire, Nucl. Phys.
{\underline{B287}, 717, (1987) }
\bibitem{Antoniadis}  I. Antoniadis, Phys. Lett. {\underline{B246}, 317 (1990)}
\bibitem{Davidson}  A. Davidson, Mod. Phys. Lett. {\underline{A13}, 2178 (1998)}
\bibitem{Nash}  J.  Nash, Annals of Maths. { \underline{63},,20 (1956)}
\bibitem{Teitelboim} C. Teitelboim,  Ann. Phys. (N.Y.) { \underline{79}, 542
(1973)}  
\bibitem{Hojman}  S. Hojman, K.  Ku\v{c}har \&  C. Teitelboim, Ann. of Phys. {
\underline{96}, 88, { (1976)}}.
\bibitem{Sorkin}  J. L. Friedman \&  R. D. Sorkin, Phys. Rev Lett. {
\underline{44}, 1100 (1990)} 
\bibitem{Janet}   M. Janet, Ann. Soc. Pol. Mat { \underline{5}, (1928)}, E. Cartan
Ibid {  \underline{6} ,1 (1927)}.
\bibitem{Greene}  R. Greene, Memoirs  Am.  Math. Soc {  \underline{97},
(1970)}. 
\bibitem{Ashtekar} A. Ashtekar, {\em Quantum Mechanics of Geometry} Preprint,
Penn. State University,  CGPG-98-12-5,  gr-qc/9901023
\bibitem{Hawking} S. Hawking, Phys. Rev {  \underline{D37}, 904  (1988)}.
\bibitem{Balachandra} A. P. Balachndran et all, Nucl. Phys.
{\underline{B446}, 299,  (1995)}.    
\bibitem{Dowker} F. Dowker \&  R. D. Skirme,  Class. Quant. Grav. {
\underline{115}, 1153, (1998)}.
\bibitem{Klein}  O. Klein,  Nature, {  \underline{118}, 516, (1926)}
\bibitem{DeWitt}   DeWitt, B.S. \& Molina-Par\'{\i}s, C.  Mod. Phys. Lett. {
\underline{A13}, 2475 (1998)}
\bibitem{Maia} M. D. Maia, Class. Quant. Grav. { \underline{6}, 173 (1989)}
\bibitem{ME} E. M. Monte \&  M. D. Maia, J. Math. Phys.   { \underline{37}, 1972 (1996) }.
\end{thebibliography}
\end{document}